\documentclass[conference]{IEEEtran}
\IEEEoverridecommandlockouts
 \usepackage[table]{xcolor}
\usepackage{caption}
\usepackage{subcaption}
\usepackage{threeparttable}
\usepackage{booktabs} 

\usepackage[letterpaper, left=0.625in, right=0.625in, top=0.72in, bottom=0.95in]{geometry}

\usepackage{cite}
\usepackage{amsmath,amssymb,amsfonts}
\usepackage{pifont}  
\usepackage{booktabs} 
\usepackage{tikz}
\usetikzlibrary{positioning, arrows.meta, shapes.geometric}
\usepackage{graphicx}
\usepackage{textcomp}
\usepackage{algorithm}
\usepackage{color,xcolor}
\usepackage{listings}
\lstdefinestyle{toolbox}{
  basicstyle=\footnotesize\ttfamily,
  frame=single,
  framerule=0.5pt,
  rulecolor=\color{gray!55},
  backgroundcolor=\color{gray!6},
  xleftmargin=10pt,
  xrightmargin=4pt,
  framexleftmargin=6pt,
  framexrightmargin=6pt,
  framextopmargin=4pt,
  framexbottommargin=4pt,
  aboveskip=8pt,
  belowskip=8pt,
  breaklines=true,
  columns=fullflexible,
  keepspaces=true,
  showstringspaces=false,
  commentstyle=\itshape\color{gray!60!black},
  keywordstyle=\bfseries\color{blue!55!black},
  morekeywords={for,each,if,then,else,return,in,not,and,or,fetch,emit,run,parse,break,procedure,function},
}
\usepackage{url}
\usepackage{multirow}
\usepackage{subfig}
\usepackage{xspace}
\usepackage{tablefootnote}
\usepackage{mathtools}
\usepackage{tabularx}
\usepackage{algpseudocode}
\usepackage{float}
\usepackage{framed}
\usepackage{fancyhdr}
\usepackage[numbers]{natbib}
\usepackage{seqsplit}  

\newcommand{\anoncite}[1]{[*]}
\usepackage{ulem}
\def\BibTeX{{\rm B\kern-.05em{\sc i\kern-.025em b}\kern-.08em
    T\kern-.1667em\lower.7ex\hbox{E}\kern-.125emX}}
\usepackage[T1]{fontenc}     
\usepackage{mathptmx} 
\begin{document}

\title{SADE: Symptom-Aware Diagnostic Escalation for LLM-Based Network Troubleshooting\\
}


\author{Kuan-Hao Tseng*, Niruth Bogahawatta*, Yasod Ginige*, Kosta Dekic*, Arunan Sivanathan\textsuperscript{\dag}, Suranga Seneviratne*\\
*\textit{University of Sydney, Sydney, Australia}\\
\textsuperscript{\dag}\textit{University of New South Wales, Sydney, Australia}\\
Email: ktse0346@uni.sydney.edu.au, niruth.savin@sydney.edu.au, a.sivanathan@unsw.edu.au,\\
\{firstname.lastname\}@sydney.edu.au}


\maketitle

\begin{abstract}

Large language model (LLM) agents are increasingly applied to network
troubleshooting, but root-cause localization on public benchmarks remains
well below practical deployment thresholds. We argue this is because
existing agents do not encode the disciplined, layer-by-layer methodology
that human network engineers use, and instead rely on free-form
deliberation that conflates evidence acquisition with hypothesis
commitment. We present \textbf{SADE} (Symptom-Aware Diagnostic
Escalation), an agent that encodes the classical Cisco troubleshooting
methodology as an explicit policy. SADE pairs a phase-gated diagnostic
workflow, which separates evidence acquisition from hypothesis commitment,
with a routed library of fault-family skills and high-yield diagnostic
helpers. On a held-out 523-incident slice of the public NIKA benchmark
covering eleven unseen scenarios, SADE improves root-cause F1 by
$37$ percentage points over a ReAct + GPT-5 baseline; a model-controlled
comparison against the same Claude Sonnet backend without the SADE
policy attributes $22$ of those points to the diagnostic policy alone,
showing that the gain is not a side-effect of the model upgrade.

\end{abstract}


\section{Introduction}
\label{sec:Introduction}


Modern computer networks fail in ways that are both costly and slow to diagnose. For instance, a single BGP misconfiguration in 2021 caused a 6-hour outage across Meta platforms, affecting billions of users~\cite{meta_outage_2021}. Similar incidents have occurred at telcos and cloud providers, affecting not only consumers and applications but also critical emergency services~\cite{crtc_rogers_2024,aws_dynamodb_2025}. In each case, operator or automation errors triggered the outages, but recovery was slow: network engineers had to diagnose cascading, multi-layered symptoms under pressure, often while their own monitoring and remote-access tools were also impaired. The cost of a network fault is dominated by how long it takes operators to identify the root cause.

Network operational complexity has driven automation efforts, such as Intent-based networking and zero-touch service management, aiming to reduce operator burden by translating high-level goals into low-level configurations and policy-driven workflows~\cite{etsi_zsm_2018, ibn_overview_minhas2024}. With their advent, LLMs (Large Language Models) have also been applied to networking tasks, including intent translation, configuration generation, and network management assistance~\cite{llmibn, llnet, llmnetcfg, netconfeval, netllm, survey}. While important, these efforts primarily address specifying, generating, or managing configurations during normal operation. They do not fully address diagnosing an already-failing network with incomplete, cascading symptoms. An effective diagnostic system must decide what evidence to collect next, inspect the live network, update its hypothesis as observations arrive, and localize the root cause under time and query constraints. This gap motivates a shift to agentic network fault diagnosis, where the challenge is not producing configurations from intent, but actively gathering evidence and reasoning toward a root cause.

\begin{figure}[t]
    \centering
    \includegraphics[width=\linewidth]{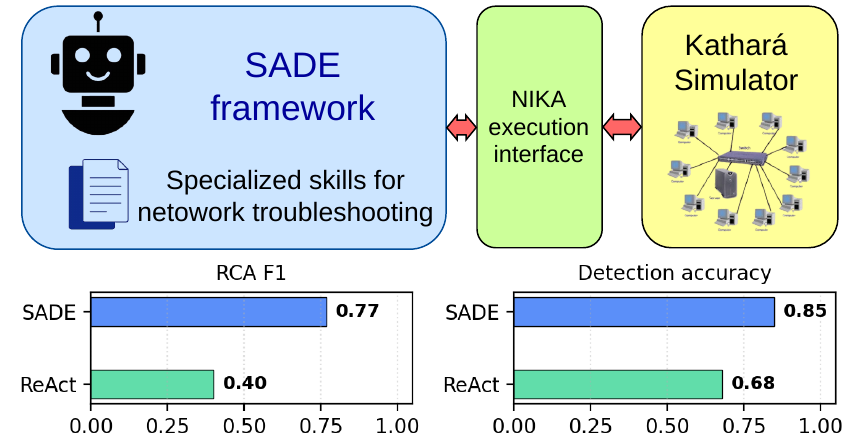}
    \caption{SADE system overview. An LLM agent with specialized troubleshooting skills interacts with Kathara-emulated topologies  via the NIKA interface, achieving a root-cause F1 score of $0.77$ compared to $0.40$ achieved by ReAct.}
    \label{fig:sade_architecturehigh}
    \vspace{-3mm}
\end{figure}

Recent work has explored the use of LLMs for network troubleshooting. NetAssistant~\cite{wang2024netassistant} uses natural-language queries to select pre-authored workflows in data-center networks, while BiAn~\cite{bian2025sigcomm} applies a hierarchical LLM pipeline with early-stop criteria for failure localization. The NIKA benchmark~\cite{wang2025networkarenabenchmarkingai} uses Kathara-emulated~\cite{kathara} topologies and the Model Context Protocol (MCP)~\cite{mcp_2024} to evaluate agentic diagnosis on common network faults. While these works shift toward troubleshooting, they remain limited by predefined workflows, fixed pipelines, or basic task evaluations that do not capture how expert operators systematically narrow faults across layers. This motivates a diagnostic agent that exploits iterative evidence collection, hypothesis revision, and disciplined escalation from symptoms to root causes.
 

To this end, we propose \textbf{SADE} (Symptom-Aware Diagnostic Escalation), a methodology-grounded LLM agent for network fault diagnosis (Figure~\ref{fig:sade_architecturehigh}). SADE is based on the premise that effective troubleshooting requires more than giving an LLM access to network tools: the agent must follow a disciplined diagnostic procedure that determines what to inspect first, how to interpret symptoms, and when to escalate across layers. More specifically, we make the following contributions.



\begin{itemize}

\item We propose SADE, an LLM-based agentic framework for network fault diagnosis that follows a multi-stage troubleshooting approach comprising: (i) initial scan, (ii) deep network scan, (iii) symptom-to-fault-family matching, and (iv) network skills library-based root cause analysis.



\item We evaluate SADE on the public NIKA benchmark against two baselines: ReAct + GPT-5 and Claude Code. SADE achieves a root cause analysis (RCA) F1 score of 0.77, outperforming the NIKA-native ReAct + GPT-5 baseline (0.40) and Claude Code (0.55), while requiring fewer diagnostic steps. These results demonstrate improved symptom confirmation, more accurate fault-family skill selection, and reduced unnecessary tool use.

\item We further demonstrate that SADE can troubleshoot network faults across topologies of varying sizes, including those exceeding 100 nodes, owing to its cross-stack diagnostic capabilities. Consequently, SADE achieves a task completion rate approximately 5\% higher than that of the best-performing baseline.




\item We release an anonymous GitHub repository containing the SADE source code, diagnostic skill library, and experiment utility scripts to support reproducibility and facilitate future research.\footnote{\textcolor{blue}{https://github.com/Overlxrd-uwu/SADE-NetworkAgent}}

\end{itemize}



The rest of the paper is organized as follows. Section~\ref{sec:RelatedWork} reviews related work. Section~\ref{sec:method} presents the SADE design. Section~\ref{sec:evaluation} describes our experimental setup. Section~\ref{sec:results} reports results and ablations. Section~\ref{sec:discussion} discusses limitations and concludes.

\section{Related Work}
\label{sec:RelatedWork}

Before the emergence of LLMs, network automation efforts focused on configuration and policy management. Intent-based networking~\cite{ibn_overview_minhas2024} and zero-touch service management~\cite{etsi_zsm_2018} translate high-level objectives into configurations and workflows. Static analysis tools like Batfish~\cite{batfish} verify configuration properties before deployment. Learning-based systems leverage historical data and graph neural networks for configuration recommendation~\cite{configreco, netgenius} or mobile-network tasks~\cite{mobilenetgnn}. While these approaches help specify goals and validate configurations, they do not address open-ended fault diagnosis in live, degraded networks, where troubleshooting requires collecting evidence, invoking tools, and revising hypotheses as observations accumulate. LLMs enable new diagnostic approaches by reasoning over natural language, configurations, tool outputs, and heterogeneous operational evidence.

\subsection{LLM-Based Network Configuration Management}


\label{sec:rw:llmintenttranslation}

LLNet~\cite{llnet} demonstrates that even a small language model, paired with an intermediate JSON representation, can compile natural-language intents into deployable programs across heterogeneous SDN data and control planes. Mekrache and Ksentini~\cite{llmibn} adopt the same paradigm for a 5G testbed, with explicit extensions toward 6G, embedding the LLM into an end-to-end intent lifecycle spanning decomposition, translation, and activation. Lira et~al.~\cite{llmnetcfg} situate the approach within zero-touch service management, while GeNet~\cite{genet} broadens the input modality by using a multimodal LLM to interpret topology diagrams alongside textual intent. At production scale, Confucius~\cite{confucius2025} reports the experience of deploying a multi-agent management framework at Meta, modelling workflows as directed acyclic graphs over specialised agents and gating mission-critical actions through validation and human approval. NetLLM~\cite{netllm} adapts general-purpose models to networking prediction through lightweight fine-tuning, while NetConfEval~\cite{netconfeval} benchmarks configuration-translation performance across model families and intent complexities.  Recent surveys consolidate this emerging direction and identify open challenges~\cite{survey, chatnet}. What unifies this work is that the desired outcome is known upfront and the LLM translates intent into correct configurations. In contrast, SADE targets the inverse problem: troubleshooting with an unknown cause, where the agent must probe the network interactively under incomplete evidence.

\subsection{LLMs for Network Troubleshooting}
\label{sec:rw:llm-troubleshooting}

NetAssistant~\cite{wang2024netassistant} routes natural-language diagnosis queries to pre-authored workflows in data-center networks and has been deployed in production. BiAn~\cite{bian2025sigcomm} applies a hierarchical LLM pipeline to production-scale failure localization with early-stop criteria. Donadel et~al.~\cite{donadel2024llm} evaluate zero-shot LLMs on Kathara-emulated topologies. Most directly related, the NIKA benchmark~\cite{wang2025networkarenabenchmarkingai} provides a public platform for evaluating agentic LLMs on classical network troubleshooting tasks. The majority of existing LLM-based root-cause-analysis literature~\cite{rcacopilot2024,rcagent2024,roy2024llmrca} focuses on cloud and microservice environments rather than L2/L3 networks. While these systems demonstrate the promise of LLM-assisted diagnosis, they do not explicitly address how to structure an agent's diagnostic process to move from surface-level symptoms to deeper fault reasoning when initial observations are insufficient.

\subsection{Tool-Using Agents and Reusable Skill Libraries}

Two recent general-purpose agentic paradigms inform the design of SADE. The ReAct paradigm~\cite{yao2022react} interleaves reasoning, tool invocation, and observation in a single decision loop and now underpins most modern tool-using agents, including those deployed in networking. Voyager~\cite{wang2023voyager} introduced the idea of incrementally constructing a reusable skill library from past experience, allowing an agent to accumulate procedural competence across episodes rather than re-deriving it from scratch. More recently, Anthropic's Agent Skills convention~\cite{anthropic_skills_2025} operationalises this idea for production deployment: reusable procedural knowledge is authored as \texttt{SKILL.md} files and dispatched through a central index, so that only the skills relevant to the current task are loaded into the agent's working context. SADE builds on these foundations but targets structured network troubleshooting: it separates general diagnostic methodology from fault-specific expertise, enforces their interaction through multi-stage escalation, and demonstrates on NIKA that this yields more reliable root-cause localisation than generic tool-using approaches.

\section{SADE Framework}
\label{sec:method}

\begin{figure}[t]
    \centering
    \includegraphics[width=\linewidth]{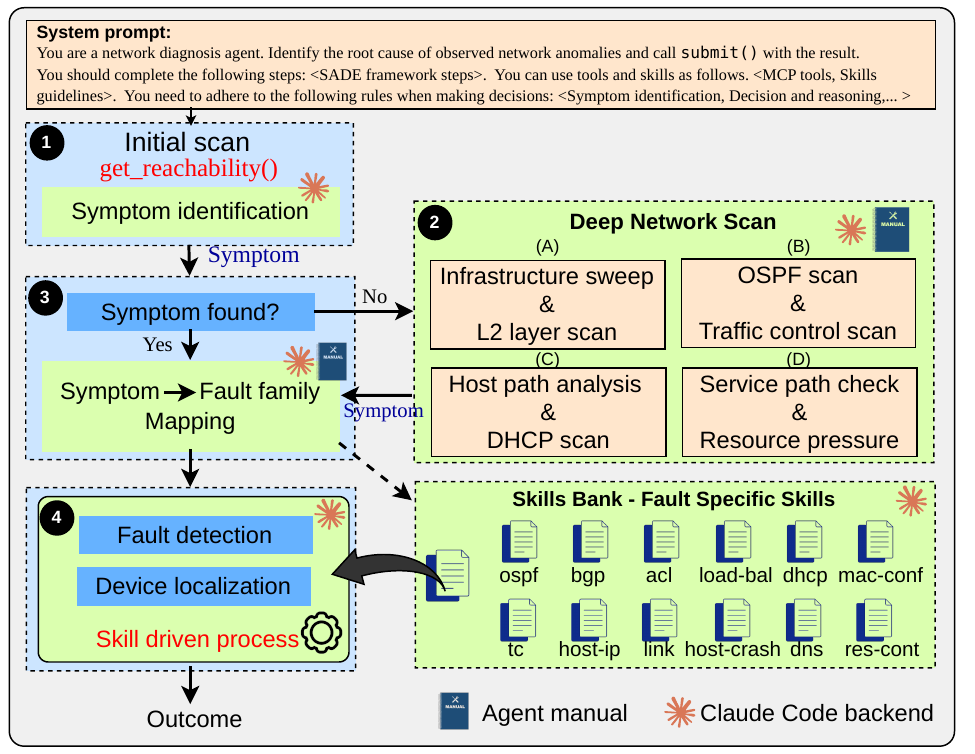}
    \caption{SADE framework. Actions that interact with the Kathar\'a simulation are marked in red text.}
    \label{fig:sade_architecture}
\end{figure}

The \textbf{SADE} (Symptom-Aware Diagnostic Escalation) framework follows the Cisco network troubleshooting methodology~\cite{cisco_troubleshooting_methods} by encoding it into an agent-driven process through agent policy. As illustrated in Figure~\ref{fig:sade_architecture}, the framework comprises four primary stages: (i) an initial scan that performs a basic reachability test to identify observable symptoms; (ii) a deep network scan that inspects the system from the data-link layer up to the application layer in four ordered phases (L2/infrastructure $\rightarrow$ control plane $\rightarrow$ host-local $\rightarrow$ service), when the initial scan fails to reveal issues; (iii) a symptom-to-fault-family mapping stage that associates detected symptoms with high-level fault categories; and (iv) a Claude skill-driven fault detection and localization process. 

The Deep Network Scan uses multiple Python helper scripts to perform specific scans to diagnose issues in different layers of the network, including infrastructure-level checks (e.g., routing, ARP, and ACL inspection) and Layer 2 anomaly detection, such as identifying duplicate MAC addresses. The identified symptoms are first mapped to the appropriate fault family using the Fault-Index, after which the Skill Bank, consisting of  15 Claude Skills, carries out the subsequent in-depth inspection to pinpoint the exact fault and accurately localize the responsible device through systematic, evidence-based reasoning. To perform the exact fault detection and localization, we use a Skills Bank that contains 15 Claude Skills, which guide the agent through systematic fault detection and localization. When executing actions on the simulator, SADE invokes the MCP tools exposed by the NIKA execution interface, which mediates all communication with the underlying Kathar\'a simulator. The experiment setup is further discussed in Section~\ref{sec:evaluation}. Below, we discuss each SADE step in detail.

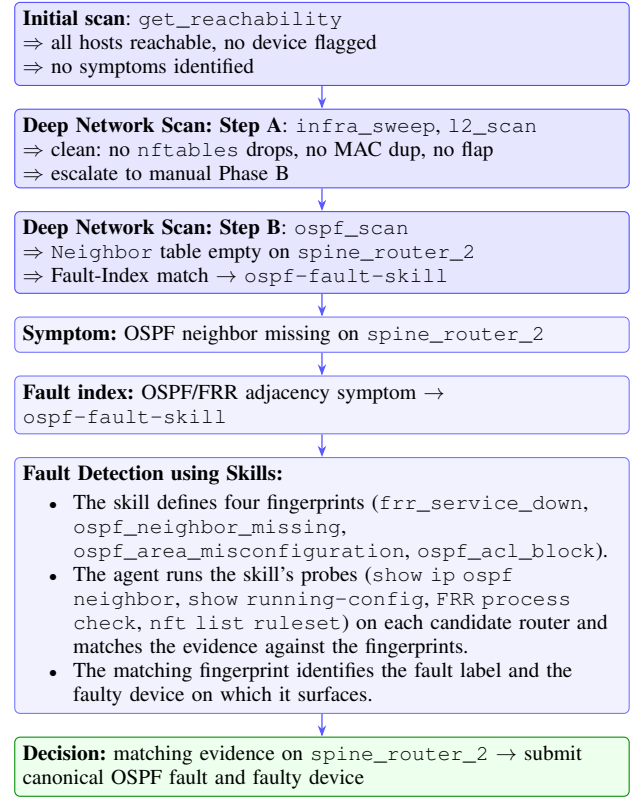
\begin{figure}[t]
\centering
\footnotesize
\begin{tikzpicture}[
  node distance=3mm,
  block/.style={
    rectangle, rounded corners=2pt,
    draw=blue!55, fill=blue!5,
    text width=0.88\linewidth,
    inner sep=3pt,
    align=left
  },
  stage/.style={
    block,
    fill=blue!10,
    draw=blue!70
  },
  highlight/.style={
    block,
    fill=green!7,
    draw=green!55!black
  },
  arr/.style={-{Stealth[length=1.5mm]}, blue!65, line width=0.5pt}
]

\node[stage] (s1) {
\textbf{Initial scan}: \texttt{get\_reachability}\\
$\Rightarrow$ all hosts reachable, no device flagged\\
$\Rightarrow$ no symptoms identified
};

\node[stage, below=of s1] (s2) {
\textbf{Deep Network Scan: Step A}: \texttt{infra\_sweep}, \texttt{l2\_scan}\\
$\Rightarrow$ clean: no \texttt{nftables} drops, no MAC dup, no flap\\
$\Rightarrow$ escalate to manual Phase B
};

\node[stage, below=of s2] (s3) {
\textbf{Deep Network Scan: Step B}: \texttt{ospf\_scan}\\
$\Rightarrow$ \texttt{Neighbor} table empty on \texttt{spine\_router\_2}\\
$\Rightarrow$ Fault-Index match $\to$ \texttt{ospf-fault-skill}
};

\node[block, below=of s3] (symptom) {
\textbf{Symptom:} OSPF neighbor missing on \texttt{spine\_router\_2}
};

\node[block, below=of symptom] (index) {
\textbf{Fault index:} OSPF/FRR adjacency symptom $\rightarrow$ \texttt{ospf-fault-skill}
};

\node[block, below=of index] (skill) {
\textbf{Fault Detection using Skills:}\\

\begin{itemize}
    \item The skill defines four fingerprints (\texttt{frr\_service\_down}, \texttt{ospf\_neighbor\_missing}, \texttt{ospf\_area\_misconfiguration}, \texttt{ospf\_acl\_block}).

    \item The agent runs the skill's probes (\texttt{show ip ospf neighbor}, \texttt{show running-config}, \texttt{FRR process check}, \texttt{nft list ruleset}) on each candidate router and matches the evidence against the fingerprints.

    \item The matching fingerprint identifies the fault label and the faulty device on which it surfaces.
\end{itemize}

};

\node[highlight, below=of skill] (submit) {
\textbf{Decision:} matching evidence on \texttt{spine\_router\_2}
$\rightarrow$ submit canonical OSPF fault and faulty device
};

\draw[arr] (s1) -- (s2);
\draw[arr] (s2) -- (s3);
\draw[arr] (s3) -- (symptom);
\draw[arr] (symptom) -- (index);
\draw[arr] (index) -- (skill);
\draw[arr] (skill) -- (submit);

\end{tikzpicture}
\caption{Example workflow for troubleshooting an OSPF error.}
\label{fig:skill_fetch_example}
\end{figure}

\subsection{Deep Network Scan}\label{subsec:deep_scan}

\subsection{Initial Scan}\label{subsec:initial_scan}

The initial scan stage calls \texttt{get\_reachability()}, a prebuilt MCP function from NIKA to scan the network Kathar\'a simulator. It returns the complete ping results from each host to all other hosts in the network, returning a matrix of all-pairs. Each reachability entry records the source, destination, transmitted and received packets, loss percentage, status, and destination IP information. SADE uses this output to determine whether a genuine symptom is immediately observable. This classification is performed based on criteria defined in SADE’s system prompt: entries with non-zero packet loss, timeouts, refused connections, or ICMP unreachable responses are treated as candidate symptoms and forwarded to Step~3, while all other cases proceed to Deep Network Scan (Step 2). We note that \texttt{get\_reachability()} can produce ambiguous outputs when name resolution fails at the source, resulting in rows with \texttt{$status=``unknown"$} and null transmission. This typically occurs due to missing or incorrect DNS entries, even if the underlying L3 path is functional. To resolve this ambiguity, SADE performs a direct-IP probe by re-running \texttt{ping} from the same source using the destination IP provided in the original \texttt{get\_reachability} output. By bypassing DNS, this step distinguishes between a resolver issue and an actual network path failure before assigning the symptom to a fault family.

At the end of the scan, the identified symptoms are forwarded to the next stage when present; if no symptoms are detected, a “symptoms not found” message is passed instead.

In cases where the Initial Scan (Step 1)
fails to identify any fault symptoms, we conduct a systematic network scan, following the instructions in the agent manual as shown in Appendix~\ref{app:examples}. This step is necessary because we observed that many faults do not appear as simple host-to-host reachability failures. For example, routing control-plane issues can be masked by redundant paths, services can remain operational while their ports are filtered, and apparent host disconnections can be caused by underlying ARP or DNS-related behavior. As illustrated in Figure~\ref{fig:skill_fetch_example}, the scan starts with L2 and infrastructure evidence, then moves to control-plane and routing evidence, host-local evidence, and finally service and resource-pressure evidence.  Table~\ref{tab:deep_scan_helpers} summarizes the main helper scripts used in this process. An example is discussed in the Appendix~\ref{app:examples}. This bottom-up ordering follows from the observability behaviour described above and is guided by Cisco's structured troubleshooting methodology~\cite{cisco_troubleshooting_methods}. Because lower-layer faults often surface as higher-layer symptoms, the scan first confirms L2 and routing evidence, so a healthy-looking service or host reading is not accepted while an underlying ACL, routing, or ARP fault remains uninspected. 


The identified symptoms will be passed to Step 3 to map to a fault family.

\begin{table*}[!t]
\centering
\footnotesize
\renewcommand{\arraystretch}{1.15}
\setlength{\tabcolsep}{4pt}
\rowcolors{2}{gray!6}{white}
\caption{Helper scripts used in SADE's deep network scan.}
\label{tab:deep_scan_helpers}
\begin{tabularx}{\linewidth}{@{}>{\raggedright\arraybackslash}p{0.16\linewidth} >{\raggedright\arraybackslash}p{0.22\linewidth} >{\raggedright\arraybackslash}X@{}}
\toprule
\textbf{Scan phase} & \textbf{Helper script} & \textbf{Functionality / output} \\
\midrule

L2 / infrastructure &
\seqsplit{\texttt{infra\_sweep.py}} &
Runs \texttt{nft list ruleset}, \texttt{ip -br addr}, \texttt{ip route}, \texttt{arp -n}, and \texttt{/etc/resolv.conf} on every device. Returns flagged devices and the failure: firewall drop, missing IP / route / gateway, ARP-cache mismatch, or broken resolver. \\

L2 / infrastructure &
\seqsplit{\texttt{l2\_snapshot.py}} &
Reads interface state, MAC, and bridge membership on every device. Returns any interface pair sharing a \texttt{link/ether} address (duplicate-MAC fault, invisible to reachability checks). \\

Control plane / routing &
\seqsplit{\texttt{ospf\_snapshot.py}}, \seqsplit{\texttt{bgp\_snapshot.py}} &
Queries the FRR (FRRouting) daemon on every router for OSPF / BGP state. Returns per router: daemon status, advertised networks, neighbor adjacency, route counts, and a convergence verdict. \\

Traffic behavior &
\seqsplit{\texttt{tc\_snapshot.py}} &
Runs \texttt{tc qdisc show} on every active interface. Returns only interfaces with a non-default qdisc (shaping or rate-limiting), each with a \texttt{tc -s} stats summary. \\

Host-local &
\seqsplit{\texttt{host\_path\_snapshot.py}}, \seqsplit{\texttt{dhcp\_link\_history.py}}, \seqsplit{\texttt{safe\_reachability.py}} &
On a chosen host, checks the kernel path to the target (\texttt{ip route get} + next-hop ARP), parses recent DHCP / link events from logs, and runs a fallback reachability sweep when MCP \texttt{get\_reachability()} fails. Returns the outbound path, link-flap history, and a partial reachability matrix. \\

Service / resource &
\seqsplit{\texttt{service\_snapshot.py}}, \seqsplit{\texttt{pressure\_sweep.py}} &
Checks DNS, hostname resolution, per-URL HTTP timing on the host; samples CPU and socket spikes, daemon presence, and stress processes on every device. Returns service-side health and per-device contention signals separating load-induced from configuration faults. \\

\bottomrule
\end{tabularx}
\vspace{-3mm}
\end{table*}

\subsection{Symptom to Fault Family Mapping}

SADE uses a fault index that maps the found symptoms to high-level fault categories (fault families) using a well-defined rule set and then decides the fault-relevant skill to be used for further exploration from the Skills Bank. The agent uses the rule set and the examples given to map the symptoms to the fault family and the skill through logical reasoning. The index also includes disambiguation logic for cases where a single symptom may originate from multiple layers.

For example (Figure~\ref{fig:skill_fetch_example}), an OSPF neighbor missing on a router could indicate either a link fault or a routing-layer fault (where the link is up but the control plane is broken). SADE's index disambiguates by first checking the interface state via \texttt{infra\_sweep}: if the interface is DOWN, the symptom routes to \texttt{link-fault-skill}, which works through its checklist and escalates if the pattern is not resolved there; if the interface is UP and the OSPF neighbor table is vacant, the decision is routed to \texttt{ospf-fault-skill}, which narrows down the possible root cause among its fingerprints. The idea is that the symptom index does not statically map a symptom to a family --- it routes the agent only after evidence on the suspected layer supports the choice.


\subsection{Fault Detection and Localization}\label{subsec: faulty_detection}


After the fault family is selected, SADE fetches the corresponding fault-specific Claude skill using the \texttt{Skill} tool. Each skill is a declarative \texttt{SKILL.md} document --- written in Markdown and supplied to the agent as in-context instructions at fetch time --- that names the canonical failure modes in that family, the leading signals that distinguish them, the exact probes used to confirm each one, helper scripts, guardrails, and stop conditions. The mapping from probe output to fingerprint is performed by the agent at run time by reasoning over the returned evidence against the skill's textual descriptions, rather than by a fixed dispatch table, so the same skill applies to any topology and any device on which the family's symptom surfaces. We discuss the content of a skill file further in the Appendix~\ref{app:skills_overview}. The agent then executes the probes specified by the selected skill and compares the collected evidence against the skill’s fault fingerprints. Here, a probe is a targeted shell command or MCP call, such as a firewall-rule listing via \texttt{nft}, an OSPF-neighbour query through \texttt{vtysh}, or a daemon-presence check via \texttt{pgrep}, designed to gather structured evidence from a suspected device, allowing direct matching against the skill’s description. Once a fingerprint matches, the faulty device is identified as the one on which the decisive probe yields a positive fingerprint. The system then submits the corresponding canonical fault label, defined by the fingerprint itself, as the final result. For instance, in Figure~\ref{fig:skill_fetch_example}, the discovered symptom is missing OSPF neighbour, and it is mapped to the OSPF fault skill, whose four fingerprints and matching probes confirm the affected spine router as the faulty device. We present the algorithmic summarization of the SADE agent in Algorithm~\ref{alg:sade} in the Appendix.

\section{Experiment Settings}
\label{sec:evaluation}

We evaluate SADE on the public NIKA benchmark~\cite{wang2025networkarenabenchmarkingai}. The evaluation compares SADE against two baselines: the published ReAct baseline from NIKA and Anthropic's recent Claude Code agent, which has demonstrated excellent command-line and automation skills. 


\subsection{Experimental Setup}
\label{sec:setup}

\paragraph{NIKA dataset and evaluation split}

NIKA is built on Kathar\'a~\cite{kathara}, a Docker container-based network emulator, and exposes each topology to an LLM agent through the Model Context Protocol (MCP)~\cite{mcp_2024}. The dataset contains $640$ incidents covering $54$ network issues across 12 scenario configurations, grouped into five network scenarios: data-centre CLOS, three-tier campus, ISP backbone, SDN, and P4. Each scalable scenario configuration is instantiated at up to three topology sizes: small, medium, and large.

Since NIKA does not provide a predefined training/test split, we reserve one scenario configuration, \texttt{ospf\_dhcp}, for SADE development (training set). We choose this configuration because it appears at all three topology sizes and covers $36$ of the $54$ network issues, yielding $108$ development incidents across the three topology sizes. The remaining four network scenarios contain $532$ held-out incidents across topology sizes and network issues. During execution, however, nine of these incidents were skipped at deployment time by the NIKA injector. We therefore report results on the $523$ held-out incidents for which all three systems produced complete runs, ensuring that SADE, CC-Baseline, and ReAct+GPT-5 are compared on the same workload. Table~\ref{tab:test_breakdown} summarizes these $523$ matched test cases, broken down by scenario configuration, topology size, and NIKA root-cause category.


\begin{table}[!t]
\centering
\scriptsize
\setlength{\tabcolsep}{2pt}
\renewcommand{\arraystretch}{1.05}
\caption{Test-set composition ($N=523$ matched triples).}
\label{tab:test_breakdown}

\textit{By scenario $\times$ topology size}
\vspace{1pt}

\begin{tabular}{@{}lrrrrrrrr@{}}
\toprule
\textbf{Size} &
\textbf{\texttt{dc\_clos}} &
\textbf{\texttt{dc\_clos}} &
\textbf{\texttt{ospf\_ent.}} &
\textbf{\texttt{rip\_small}} &
\textbf{\texttt{sdn}} &
\textbf{\texttt{sdn}} &
\textbf{P4} &
\textbf{Total} \\
&
\textbf{\texttt{\_bgp}} &
\textbf{\texttt{\_service}} &
\textbf{\texttt{\_static}} &
\textbf{\texttt{\_int\_vpn}} &
\textbf{\texttt{\_clos}} &
\textbf{\texttt{\_star}} & & \\
\midrule
s      & 22 & 33 & 26 & 23 & 18 & 19 & --  & 141 \\
m      & 22 & 33 & 26 & 23 & 19 & 19 & --  & 142 \\
l      & 22 & 33 & 26 & 24 & 19 & 18 & --  & 142 \\
single & -- & -- & -- & -- & -- & -- & 98  & 98  \\
\midrule
\textbf{Total} & 66 & 99 & 78 & 70 & 56 & 56 & 98 & \textbf{523} \\
\bottomrule
\end{tabular}

\vspace{6pt}
\textit{By NIKA root-cause category}
\vspace{1pt}
\begin{tabular}{@{}rrrrrrr@{}}
\toprule
\textbf{End-host} & \textbf{Misconfig.} & \textbf{Link fail.} & \textbf{Resource cont.} & \textbf{Node error} & \textbf{Under attack} & \textbf{Total} \\
\midrule
124 & 123 & 99 & 82 & 60 & 35 & \textbf{523} \\
\bottomrule
\end{tabular}
\vspace{-5mm}
\end{table}

\paragraph{SADE implementation and Baselines}

SADE is implemented as an agentic diagnostic system on top of the Claude Agent SDK, using Claude Sonnet~4.6~\cite{anthropic_sonnet46_2025} as the underlying LLM. The implementation consists of a four-step method, described in Section~\ref{sec:method}.


For each test incident, SADE interacts with the topology only through the MCP tools provided by NIKA. The agent is given a fixed budget of 20 API turns, after which it must submit its diagnosis in the NIKA submission format. A turn is one LLM API call; the agent may issue multiple (parallel) tool calls within a single turn before yielding back to the runner. We adopt the $20$-step cap from NIKA's released ReAct agent (its LangGraph \texttt{recursion\_limit}), applied as a $20$-turn cap on SADE and CC-Baseline so all three systems share the same per-incident reasoning budget.

For comparison, we use two baselines. The first is NIKA's published ReAct baseline, which uses a two-agent LangGraph workflow: one agent investigates the incident and another produces the final submission. We use the strongest configuration reported in NIKA, based on GPT-5~\cite{openai_gpt5_2025}. The second baseline is a Claude Code agent using Claude Sonnet~4.6, the same model backend as SADE. This baseline does not use SADE's extra features such as the skill library or helper scripts, and therefore measures how much of SADE's performance comes from the proposed diagnostic design rather than from the underlying Claude model alone. In our comparison, all systems are run on the same 523 test incidents, use the same 20-turn limit, and submit their answers in the NIKA submission format.

\subsection{Evaluation Metrics}

We use the same evaluation measures as NIKA. Detection performance is measured by accuracy on the binary \texttt{is\_anomaly} field, which indicates whether the agent correctly identifies the presence of an anomaly. Root-cause performance is measured using F1 over the set of predicted root-cause labels, since an incident may involve more than one relevant label. In addition to these automatic metrics, we use NIKA's LLM-as-judge protocol with \texttt{gpt-5-mini}~\cite{openai_gpt5_2025}. The judge assigns scores from $1$ to $5$ across five rubric axes, relevance (localization), correctness (diagnostic actions), efficiency (tool usage), clarity (reasoning trace), and final outcome (match against ground truth), and also returns an aggregate overall score based on the judge's own reasoning over these axes. If an agent does not produce a valid submission within the turn limit, we treat the incident as an unsuccessful diagnosis and assign $\text{F1}=0$ for root-cause evaluation. To measure efficiency, we also record input and tokens, and the number of tool calls made during each run.

\section{Results}
\label{sec:results}

\begin{figure}[h!]
    \centering
    \includegraphics[width=0.9\linewidth]{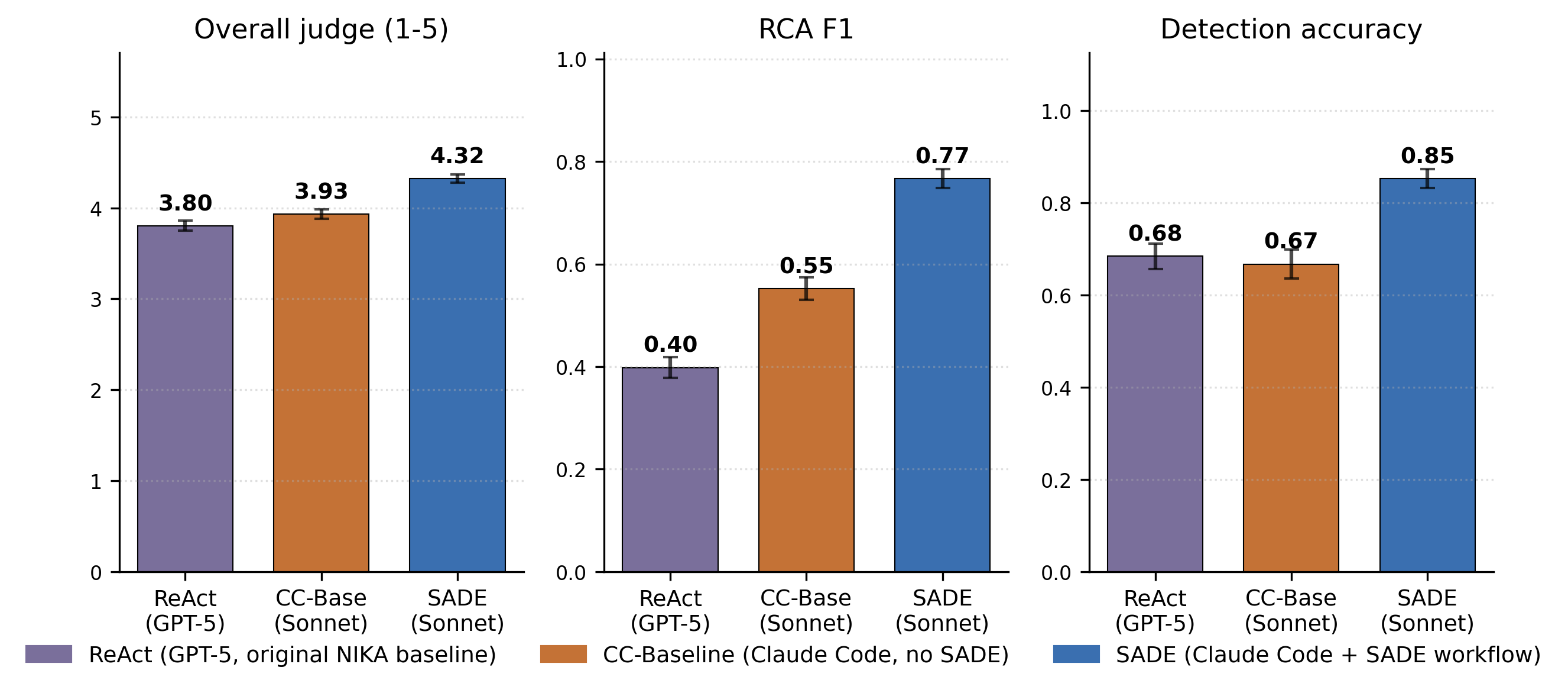}
    \caption{Overall score, RCA F1, and detection accuracy on the matched test cases}
    \label{fig:headline}
    \vspace*{-1.1\baselineskip}
\end{figure}

This section evaluates SADE against two baselines on the NIKA fault-diagnosis benchmark: ReAct with GPT-5 and a Claude-Code baseline using the same Claude Sonnet 4.6 backbone as SADE. We first report overall diagnostic correctness on 523 test cases, then examine why the gains arise by analyzing cross-stack diagnostic behavior, submission reliability, tool-call efficiency, and robustness across topology sizes.

Figure~\ref{fig:headline} reports the main results on the 523 matched NIKA test cases, where each case contains the same fault type, network scenario, and topology size across all three systems. SADE achieves the highest score on every reported correctness metric. Its mean overall judge score is 4.32, compared with 3.93 for the Claude-Code baseline and 3.80 for ReAct. SADE also improves diagnostic-specific metrics, reaching an RCA F1 of 0.77 compared to 0.55 for the same-backbone Claude-Code baseline and 0.44 for ReAct, and a detection accuracy of 0.85 compared to 0.67 and 0.68, respectively. Since SADE and the Claude-Code baseline use the same Claude Sonnet backbone, this gap shows that the improvement is not due to the model alone, but to SADE’s structured diagnostic workflow. These results support the central claim of the framework: guiding the agent through symptom-driven evidence collection improves both anomaly detection and root-cause identification, particularly in cases where the observed symptoms span multiple layers of the network stack.


Next, we examine which diagnostic cases benefit most from this structured workflow.

\subsection{Cross-stack Diagnostics}
\label{subsec:cross_stack_diagnostics}

To make the multi-layer nature of these cases concrete, we first focus on cross-stack diagnostics: faults where the visible symptom appears in one part of the network, but the root cause is exposed only by checking another part of the stack. In such cases, no single observation is sufficient. Reachability, routing state, filtering rules, host state, and service behavior may each provide partial evidence, but the diagnosis becomes clear only when these observations are connected. Table~\ref{tab:cross_stack_examples} summarizes representative examples from the matched NIKA test cases. These examples are a subset of the evaluated cases. For each case, the table reports the fault family, scenario/topology, ambiguity introduced by the fault, and diagnostic checks that expose the root cause. The corresponding ground truth, agent submissions, judge scores, and trace evidence are provided in Appendix~\ref{app:case_traces}.

The cases in Table~\ref{tab:cross_stack_examples} illustrate how this ambiguity arises in practice. In \texttt{bgp\_acl\_block}, which appears in the \texttt{dc\_clos\_service/l} topology, the fault is misleading because the network is not simply unreachable. Basic reachability checks may still succeed, making the system appear healthy if diagnosis stops at whether hosts or routers respond. The affected traffic, however, is more specific. BGP messages are carried over TCP/179, and in this case they are silently dropped by an ACL on a spine router. The root cause is therefore not exposed by broad connectivity tests alone. It becomes visible only by connecting evidence across layers: baseline reachability shows that the network is not entirely down, the failed BGP session shows that routing-state exchange is affected, and router-level filtering rules identify the ACL entry blocking BGP traffic. A similar structure appears in \texttt{dns\_port\_blocked}, however, at the service boundary rather than in the routing control plane. The DNS pod can appear reachable, and the DNS daemon may be running, so the failure is not explained by a crashed service or unavailable pod. DNS resolution nevertheless fails because the required traffic, TCP/UDP port 53, is filtered at the pod boundary. The service exists and the pod is reachable in a general sense, yet the specific requests needed to use the service cannot enter. Both cases show why cross-stack diagnosis is difficult: a healthy observation at one layer does not rule out a fault in the protocol, port, or filtering mechanism required by the failing operation.

\begin{figure}[!tp]
    \centering
    \includegraphics[width=0.92\linewidth]{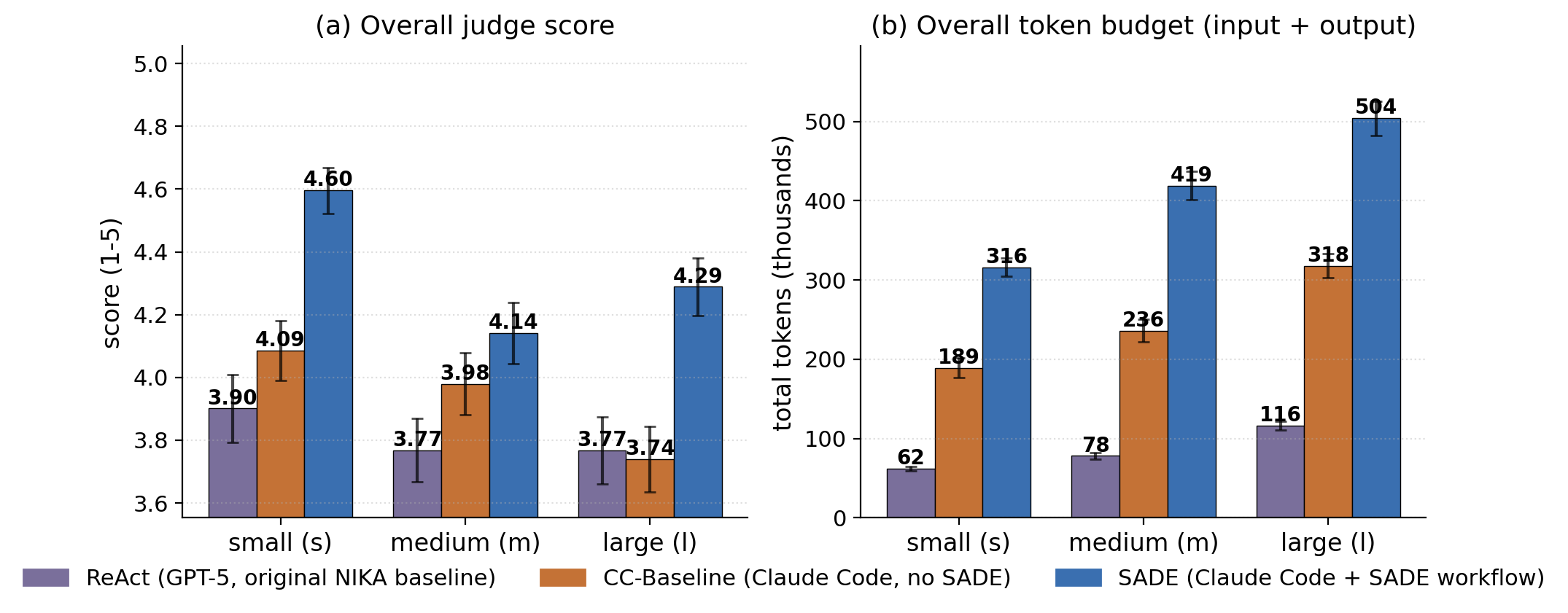}
    \caption{Sensitivity to topology size. (a) Mean overall judge score and (b) mean input-token budget by topology size, for the three sized scenario classes (averaging $11$, $27$, and $101$ nodes from \texttt{s} to \texttt{l}). Single-size (P4/SDN) labs are excluded. } 
    \label{fig:topology_scaling}
    \vspace{-4mm}
\end{figure}

The cross-stack analysis shows where SADE differs from ReAct and Claude-Code baselines. SADE succeeds because it does not treat initial reachability or service-level checks as sufficient. Instead, it continues the investigation until the symptom is connected to the relevant routing or filtering mechanism, allowing it to submit the correct root-cause family. By contrast, ReAct and Claude-Code fail because their diagnoses stop too close to the symptom: they either accept a broad but incomplete explanation, miss the protocol-specific anomaly, or fail to produce a valid final submission. The difference is therefore not simply that SADE observes more signals, but that it connects the symptom layer to the layer where the fault is introduced.

\begin{table*}[!t]
\centering
\footnotesize
\renewcommand{\arraystretch}{1.2}
\setlength{\tabcolsep}{4pt}
\rowcolors{2}{gray!6}{white}
\caption{Two cross-stack diagnostic cases from the matched test set: topology and fault label, the diagnostic ambiguity, the canonical (ground-truth) detection sequence, and the SADE traces from the session log. Both SADE runs submit the correct root cause with judge final-outcome and overall both $5/5$; comparison between baseline submissions are in Appendix~\ref{app:case_traces}.}
\label{tab:cross_stack_examples}
\begin{tabularx}{\linewidth}{@{}>{\raggedright\arraybackslash}p{0.16\linewidth} >{\raggedright\arraybackslash}X >{\raggedright\arraybackslash}X >{\raggedright\arraybackslash}X@{}}
\toprule
\textbf{Topology / Fault} & \textbf{Diagnostic ambiguity} & \textbf{Ground-truth steps} & \textbf{SADE log audit} \\
\midrule

\seqsplit{\texttt{dc\_clos\_service/l}}\newline\seqsplit{\texttt{bgp\_acl\_block}} &
Reachability looks healthy because the Clos has redundant spines, but BGP control-plane traffic is silently dropped by an ACL on one spine. &
\textbf{Step 1.}~Verify reachability.\newline
\textbf{Step 2.}~Check BGP session.\newline
\textbf{Step 3.}~List \texttt{nft} ruleset on spine.\newline
\textbf{Step 4.}~Identify drop on TCP/$179$. &
\textbf{P1.}~$70/70$ paths at $0\%$ loss; no symptom.\newline
\textbf{P4.}~\texttt{infra\_sweep} flags \texttt{spine\_router\_2\_3} alone (BGP-ACL fingerprint).\newline
\textbf{Confirm.}~\texttt{acl-skill} $\to$ \texttt{nft list ruleset}: \texttt{tcp dport/sport 179 drop} in input/forward/output chains.\newline
\textbf{Submit.}~\texttt{bgp\_acl\_block} / \texttt{spine\_router\_2\_3}; $16$ turns. \\

\seqsplit{\texttt{dc\_clos\_service/l}}\newline\seqsplit{\texttt{dns\_port\_blocked}} &
The DNS pod and \texttt{named} daemon look healthy, but TCP/UDP port $53$ is silently filtered at the pod boundary. &
\textbf{Step 1.}~Confirm DNS daemon up.\newline
\textbf{Step 2.}~Probe TCP/UDP $53$ to pod.\newline
\textbf{Step 3.}~Inspect pod \texttt{nft} ruleset.\newline
\textbf{Step 4.}~Identify filter on port $53$. &
\textbf{P1.}~All paths at $0\%$ loss; no symptom.\newline
\textbf{P4.}~\texttt{infra\_sweep} flags \texttt{dns\_pod2} (\texttt{tcp/udp dport 53 drop}).\newline
\textbf{Confirm.}~\texttt{dns-fault-skill} $\to$ \texttt{ps}+\texttt{ss}+\texttt{nft} on \texttt{dns\_pod2}: \texttt{named} (PID~$29$) listening on \texttt{10.2.0.2:53}; \texttt{nft} drops port~$53$.\newline
\textbf{Submit.}~\texttt{dns\_port\_blocked} / \texttt{dns\_pod2}. \\

\bottomrule
\end{tabularx}
\vspace{-3mm}
\end{table*}

Both rows in Table~\ref{tab:cross_stack_examples} share the same pattern: the decisive evidence is not available from the first symptom alone, and correct RCA requires tracing the failure from the visible behavior to the specific routing or filtering mechanism that explains it. This explains why SADE's gains in detection accuracy and RCA F1 are aligned with its higher judge scores on cross-stack tasks. Agents that stop at the symptom layer can miss the fault or assign it to the wrong subsystem, whereas SADE continues the diagnostic process until the trace exposes the actual root cause.


\subsection{Sensitivity to topology size}
\label{subsec:toplogy_size}

Figure~\ref{fig:topology_scaling} examines whether SADE's diagnostic advantage persists as topology size grows. NIKA contains small (\texttt{s}), medium (\texttt{m}), and large (\texttt{l}) instances of the same scenario classes, with average topology sizes of approximately 11, 27, and 101 nodes, respectively. We exclude the single-size P4/SDN cases so that the comparison reflects only scenarios that appear across multiple topology sizes.

SADE achieves the highest mean overall judge score at every topology size, indicating that its advantage is preserved as the network grows. Larger topologies contain more devices, links, routing states, and filtering rules, expanding both the set of plausible fault locations and the volume of irrelevant evidence available to the agent. The Claude-Code baseline, which uses the same Claude Sonnet backbone as SADE, declines on larger topologies, whereas SADE retains a high score across the small, medium, and large settings. This gap reflects SADE's investigation procedure rather than model capability: SADE uses observed symptoms to narrow the search to a small set of fault families and protocol layers, instead of checking the topology uniformly. This focus allows SADE to identify the correct root cause as topology size, and the number of candidate explanations grows.

A natural concern is whether this accuracy advantage on large topologies comes at the cost of an inflated diagnostic context. Figure~\ref{fig:topology_scaling}b shows otherwise. SADE consumes more tokens (input~+~output) than either baseline in absolute terms, reflecting more systematic evidence collection, but its token usage grows at a similar rate as topology size increases: SADE scales from 316k to 504k total tokens between small and large topologies ($1.6\times$), compared to $1.7\times$ for Claude-Code (189k to 318k) and $1.9\times$ for ReAct (62k to 116k). The additional tokens are therefore spent on targeted evidence collection rather than uncontrolled context growth. Together with the accuracy results in Figure~\ref{fig:topology_scaling}a, this indicates that SADE's structured workflow sustains diagnostic accuracy on larger topologies without scaling context usage faster than the baselines, supporting the same conclusion as the cross-stack analysis: SADE's gains are not confined to small or easily localized faults. 


\subsection{Tool-call Efficiency and Submission reliability}
\label{subsec:tool_call}

\begin{table}[!b]
\centering
\scriptsize
\setlength{\tabcolsep}{8pt}
\caption{Submission reliability and tool-call efficiency on the test set ($N=523$).}
\label{tab:submission_efficiency}
\begin{tabular}{@{}lrrr@{}}
\toprule
\textbf{Metric} & \textbf{ReAct} & \textbf{CC-Base} & \textbf{SADE} \\
\midrule
\multicolumn{4}{@{}l}{\textit{Submission reliability}} \\
\quad No-submissions (of 523) {\color{red}$\downarrow$}             & 46     & 80     & \textbf{22}     \\
\quad No-submission rate (\%) {\color{red}$\downarrow$}            & 8.8    & 15.3   & \textbf{4.2}    \\
\addlinespace
\multicolumn{4}{@{}l}{\textit{Tool-call efficiency}} \\
\quad Mean tool calls per session {\color{red}$\downarrow$}        & 25.4   & 26.7   & \textbf{19.9}   \\
\quad Total tool calls {\color{red}$\downarrow$}                   & 13{,}266 & 13{,}962 & \textbf{10{,}393} \\
\quad Correct submissions {\color{red}$\uparrow$}                & 268    & 271    & \textbf{382}    \\
\quad Tool calls per correct submission {\color{red}$\downarrow$}  & 49.5   & 51.5   & \textbf{27.2}   \\
\bottomrule
\end{tabular}
\vspace{-5mm}
\end{table}

The previous subsections show that SADE improves diagnostic correctness on cross-stack faults and preserves this advantage as topology size increases. We next examine whether these gains are achieved reliably within the diagnostic budget (i.e., 20 turns). Table~\ref{tab:submission_efficiency} reports two complementary measures on the test set: the rate at which each agent reaches the 20-turn budget without producing a parseable final submission, and the number of tool calls issued per correctly submitted test case. No-submission cases are distinct from incorrect diagnoses: the agent may have collected useful evidence, but failed to convert it into a valid root-cause decision. In the evaluation, these runs receive no localization or root-cause credit, so submission reliability directly affects the aggregate diagnostic scores.

SADE has the lowest no-submission rate, failing to submit in $4.2\%$ of matched sessions, compared with $8.8\%$ for ReAct and $15.3\%$ for the Claude-Code baseline. The difference is largest against the  Claude-Code baseline, where SADE reduces no-submission failures from $80$ cases to $22$ cases. This indicates that the structured diagnostic workflow does not only improve the final answer when a submission is made; it also makes the agent more likely to terminate with a valid diagnosis under the benchmark budget. The audited traces further show that the baselines fail in different ways: Claude-Code often continues investigating until the budget is exhausted without calling \texttt{submit()}, whereas ReAct more often terminates with an incorrect fault family or faulty-device set. SADE reduces the first failure mode through its Phase~3 stop-and-submit rule, which turns a matched fault-family fingerprint into an explicit termination point rather than allowing the investigation to continue indefinitely.

The tool-call rows of Table~\ref{tab:submission_efficiency} show that SADE's higher submission reliability is accompanied by lower diagnostic action cost. SADE averages $19.9$ tool calls per session, compared with $25.4$ for ReAct and $26.7$ for Claude-Code, while producing substantially more correct submissions. Normalizing by successful outcomes makes the difference clearer: SADE requires $27.2$ tool calls per correct submission, whereas ReAct and Claude-Code require $49.5$ and $51.5$, respectively. Thus, SADE does not obtain its accuracy by spending more commands within the fixed turn budget; it obtains more correct diagnoses from fewer active probes. This result is consistent with the cross-stack analysis in Section~\ref{subsec:cross_stack_diagnostics}. In those cases, the decisive step is not broad exploration, but the selection of the check that exposes the layer at which the fault mechanism occurs, such as a routing-state query, a filtering-rule inspection, a service-port probe, or a host-local configuration check. SADE's phase gates and skill-index library bias the investigation toward these decisive tests once the symptom pattern has narrowed the plausible fault families. The combined results therefore provide a cost-side explanation for the earlier accuracy gains: SADE spends fewer commands on irrelevant evidence, reaches the evidence needed for a valid submission more often, and does so without expanding the diagnostic search as topology size grows.

\section{Discussion and Concluding Remarks}
\label{sec:discussion}

We presented SADE, an LLM-based framework for network fault diagnosis that separates general diagnostic procedure from fault-family-specific expertise. SADE enforces this separation through a multi-stage workflow and a skill library that guides the agent from symptom observation to targeted evidence collection and final root-cause submission. On the NIKA benchmark, SADE achieves higher diagnostic performance than both ReAct and the Claude-Code baselines. SADE improves root-cause F1 by $36.9$ percentage points over the NIKA-native ReAct~+~GPT-5 baseline, and by $21.5$ percentage points over the same-backbone Claude-Code baseline. SADE also reduces no-submission failures by $3.6{\times}$ relative to the same-backbone baseline. Our results show that SADE’s gains are not only a consequence of the underlying language model but also of the diagnostic workflow imposed on it. Next, we discuss additional relevant findings about the NIKA benchmark, limitations of SADE, and possible future extensions.



\paragraph{Fault-injection validity}
During development, we observed that NIKA's stock injector does not always inject the intended fault for some fault families. In such cases, the benchmark label records the fault as present, but the emulated network may not contain the corresponding failure condition. This makes absolute scores difficult to interpret, because a low score may reflect either an incorrect diagnosis or an unsuccessful injection. To address this, we implemented a per-fault rescue-and-verify pipeline, \texttt{run\_nika\_break}, which checks that the intended fault is present before scoring. Table~\ref{tab:nika_limits} reports these numbers on the training set, where we manually verified each test case. Under the stock injector, SADE receives $17$ score-$1$ outcomes and $19$ score-$5$ outcomes out of the $39$ runs. After verified injection, the same set receives a score of $5$ in all $39$ cases, indicating that part of the low-score mass under stock injection is caused by failed fault installation rather than SADE's diagnostic procedure. Given the impracticality of manually verifying all 523 fault injection cases on the test set, some failures may remain undetected. However, correcting such cases would only improve the reported performance metrics.


\begin{table}[!t]
\centering
\scriptsize
\caption{NIKA injector audit: SADE plus verified injection.}
\label{tab:nika_limits}
\begin{tabular}{l@{\hspace{6pt}}r@{\hspace{6pt}}r@{\hspace{6pt}}r@{\hspace{6pt}}r@{\hspace{6pt}}r@{\hspace{6pt}}r}
\toprule
\textbf{Injection regime} & \textbf{n} & \textbf{1} & \textbf{2} & \textbf{3} & \textbf{4} & \textbf{5} \\
\midrule
Stock (\texttt{train\_obs})           & 39 & 17 & 3 & 0 & 0 & 19 \\
Verified (\texttt{manual\_injection}) & 39 &  0 & 0 & 0 & 0 & \textbf{39} \\
\bottomrule
\end{tabular}
\vspace{-3mm}
\end{table}

\vspace{2mm}
\paragraph{Skill Library and LLM Backbone} The current SADE skill index is manually authored, making it inspectable and easy to update. However, it is dependent on expert effort. Future work could explore automatic skill construction and updates from incident corpora, operator playbooks, and diagnostic history knowledge bases. In our experiments, we use Claude Sonnet 4.6 for cost reasons. Recent Claude Opus models have demonstrated increased performance in other domains, and using such a backbone would likely further improve results.


\paragraph{Per-session token budget and dollar cost}

As discussed in Section~\ref{subsec:toplogy_size} SADE consumes more tokens than the baselines. Table~\ref{tab:token_budget} reports input and output distributions separately, which are billed at different rates by Anthropic. The median token consumption for SADE is 344k input and 8k output, compared to 189k/6k for the Claude-Code baseline and 55k/10k for ReAct + GPT-5. Output is priced at $5{\times}$ the input rate, so it still contributes significantly to the cost. At publicly listed rates (GPT-5: \$1.25/\$10 per 1M input/output tokens; Claude Sonnet~4.6: \$3/\$15), an average mid-size topology session costs \$0.20 for ReAct, \$0.83 for the Claude-Code baseline, and \$1.44 for SADE. 


\begin{table}[!t]
\centering
\scriptsize
\setlength{\tabcolsep}{8pt}
\caption{Per-session token budget on the test slice ($N=523$). Input and output tokens are reported separately because they are billed at different rates. Values in thousands of tokens.}
\label{tab:token_budget}
\begin{tabular}{@{}lrrr@{}}
\toprule
\textbf{Statistic} & \textbf{ReAct (GPT-5)} & \textbf{CC-Base} & \textbf{SADE} \\
\midrule
\multicolumn{4}{@{}l}{\textit{Input tokens (k)}} \\
Min (lower whisker)  & 8   & 34  & 86  \\
Q1 (25th percentile) & 31  & 101 & 246 \\
\textbf{Median}      & \textbf{55} & \textbf{189} & \textbf{344} \\
Q3 (75th percentile) & 89  & 334 & 483 \\
Max (upper whisker)  & 173 & 659 & 838 \\
\midrule
\multicolumn{4}{@{}l}{\textit{Output tokens (k)}} \\
Min (lower whisker)  & 2   & 1   & 3   \\
Q1 (25th percentile) & 8   & 4   & 5   \\
\textbf{Median}      & \textbf{10} & \textbf{6} & \textbf{8} \\
Q3 (75th percentile) & 13  & 13  & 18  \\
Max (upper whisker)  & 22  & 26  & 36  \\
\bottomrule
\end{tabular}
\vspace{-5mm}
\end{table}

\vspace{2mm}
\paragraph{Towards live networks}
NIKA runs on Kathar'a, a container-based network emulator that provides a controlled setting for evaluating routing, filtering, host-state, and service-level faults. However, it does not capture several factors that shape production troubleshooting, including hardware forwarding behavior, NIC offloads, ASIC-level queueing, telemetry delay, and monitoring-pipeline failures. Since operational networks typically expose state through controlled telemetry and configuration systems, a deployment-ready SADE would require operator-specific wrappers over routing collectors, configuration databases, monitoring platforms, and service-observability tools. 

\vspace{40pt}
\bibliographystyle{IEEEtranS}
\normalem
\bibliography{main}

\clearpage

\appendices

\newcommand{\fp}[1]{\seqsplit{\texttt{\detokenize{#1}}}}

\lstdefinestyle{skillbox}{
  style=toolbox,
  basicstyle=\scriptsize\ttfamily,
  keywordstyle=\color{blue!55!black},
  morekeywords={for,each,fetch,emit,parse,procedure,function,if,break},
}

\section{Skill and Helper-Script Examples}
\label{app:examples}

Two examples illustrate the SADE library
(Figure~\ref{fig:examples}): a fault-family skill book
(\texttt{ospf-fault-skill}) loaded on demand when the Fault Index
routes a symptom, and a helper script (\texttt{infra\_sweep}) the
diagnosis manual runs as a Phase-A triage probe. Skill books are
declarative decision tables; helper scripts emit typed \texttt{Flag}
records keyed to fault families. The remaining 14 skills and 11 helper
scripts follow the same patterns; the full library is in the artefact
release.

\begin{figure}[H]
\centering
\begin{lstlisting}[style=skillbox]
SKILL: ospf-fault-skill
failure modes
  frr_service_down            routing daemon
                              stack not running
  ospf_neighbor_missing       router not in OSPF
                              adjacency
  ospf_area_misconfiguration  link in wrong area
                              vs. peer/topology
leading signals
  one router or routed segment loses paths
  while interfaces remain present
exact probes
  ps aux | grep 'zebra|ospfd|watchfrr'
  vtysh -c 'show ip ospf neighbor'
  vtysh -c 'show running-config'
one-pass coverage
  python scripts/ospf_snapshot.py
stop-and-submit
  when direct evidence on a router (process,
  OSPF config, or area assignment) matches
  one of the three failure modes
\end{lstlisting}
\vspace{2pt}
{\footnotesize (a) Fault-family skill book.\label{fig:skill_ospf}}

\vspace{6pt}

\begin{lstlisting}[style=skillbox]
procedure infra_sweep(lab):
  flags = []
  for each device in lab.all_devices():
    nft    = device.run("nft list ruleset")
    addr   = device.run("ip -br addr")
    route  = device.run("ip route")
    arp    = device.run("arp -n")
    resolv = device.read("/etc/resolv.conf")
    # 1. ACL fingerprint dispatch (first match wins)
    for fault, pattern in ACL_FINGERPRINTS:
      if pattern matches nft:
        emit Flag(device, fault); break
    # 2. Host-config checks
    if no default gateway in route:
      emit Flag(device, host_missing_route)
    if access iface has no IP in addr:
      emit Flag(device, host_missing_ip)
    if arp has duplicate MAC for one IP:
      emit Flag(device, mac_address_conflict)
    if resolv has malformed nameserver:
      emit Flag(device, host_incorrect_dns)
  return flags

ACL_FINGERPRINTS:
  table arp filter            -> arp_acl_block
  ip protocol ospf | proto 89 -> ospf_acl_block
  tcp dport 179               -> bgp_acl_block
  tcp dport 80                -> http_acl_block
  (tcp|udp) dport 53          -> dns_port_blocked
  icmp type                   -> icmp_acl_block
\end{lstlisting}
\vspace{2pt}
{\footnotesize (b) Phase-A helper script.\label{fig:infra_sweep}}
\caption{SADE library examples: a fault-family skill book (a) and a
Phase-A helper script (b).}
\label{fig:examples}
\end{figure}

\section{SADE Diagnosis Algorithm}
\label{app:algorithm}

Algorithm~\ref{alg:sade} formalises SADE's symptom-to-fault-family
diagnosis loop for a single incident. The agent first gathers initial
evidence via \textsc{ListProblems} and \textsc{GetReachability} and
confirms any visible symptoms (Step~1); if none are found or the
picture is ambiguous, it triggers a Step~2 deep network scan using the
helper-script set $\mathcal{H}$.

\begin{algorithm}[H]
\footnotesize
\caption{SADE diagnosis loop for one incident. $\mathcal{F}$ is the
symptom-to-fault-family index; $\mathcal{H}$ is the deep-scan
helper-script set.}
\label{alg:sade}
\begin{algorithmic}[1]
    \Require Active simulated incident $I$
    \State Load fault index $\mathcal{F}$ and helper set $\mathcal{H}$
    \State \textbf{Step 1: Initial scan}
    \State $(\mathcal{L},R) \gets \textsc{ListProblems}() \,\|\, \textsc{GetReachability}()$
    \State $S \gets \textsc{Confirm}(\textsc{ReachSymptoms}(R))$
    \If{$S$ is empty or still ambiguous}
        \State \textbf{Step 2: Deep network scan}
        \State $S \gets \textsc{Confirm}(S \cup \textsc{DeepScan}(I,\mathcal{H}))$
    \EndIf
    \If{$S$ is empty}
        \State \Return \textsc{Submit}(\texttt{is\_anomaly=False})
    \EndIf
    \While{$S$ contains an unresolved confirmed symptom}
        \State $s \gets \textsc{SelectLeadSymptom}(S)$
        \State \textbf{Step 3: Symptom-to-fault-family mapping}
        \State $f \gets \mathcal{F}[s]$
        \State \textbf{Step 4: Skill-driven detection and localization}
        \State $K_f \gets \textsc{FetchSkill}(f)$
        \State $E_f \gets \textsc{SkillProbes}(K_f,I,s)$
        \If{$E_f$ matches a fault fingerprint in $K_f$}
            \State $(\ell,D) \gets \textsc{Canonicalize}(E_f,K_f,\mathcal{L})$
            \State \Return \textsc{Submit}($\ell,D$)
        \Else
            \State $S \gets \textsc{UpdateSymptoms}(S,E_f)$; mark $s$ checked
        \EndIf
    \EndWhile
    \State \Return \textsc{Submit}(\texttt{is\_anomaly=False})
\end{algorithmic}
\end{algorithm}

Each confirmed symptom is then routed through $\mathcal{F}$ to its
owning skill (Step~3), which probes for fingerprints and submits the
canonical fault label on a full match (Step~4); otherwise the loop
updates the symptom set and continues until either a fingerprint
matches or every symptom has been checked. The negative-anomaly return
is reachable only after the deep scan completes, so a clean Phase-1
reachability snapshot never triggers a no-anomaly submission alone.


\section{Representative Case Evidence}
\label{app:case_traces}

Table~\ref{tab:cases} reports per-agent diagnostic outcomes for two
cross-stack faults, showing where SADE's structured workflow recovers
the correct root cause while both baselines fail. Case~1 is presented
as a worked example: SADE's Phase-1 reachability scan returns 70/70
paths at $0\%$ loss with no symptom, so it escalates to a Phase-2
\texttt{infra\_sweep} that flags the affected spine alone via the
BGP-ACL fingerprint, and the \texttt{acl-skill} confirms the spine
\texttt{nft} drop on TCP/179 in 16 turns.

Both failure modes reveal the same gap: lacking an explicit
symptom-to-fault-family routing step, the baselines either anchor on
the first lead and never falsify it (ReAct) or widen probes without a
stop condition (CC-B). SADE's skill index forces a commitment to one
fault family per symptom, and the skill's stop-and-submit rule
prevents the budget-exhaustion failure mode entirely.

\begin{table}[H]
\centering\scriptsize
\renewcommand{\arraystretch}{1.1}
\setlength{\tabcolsep}{3pt}
\caption{Two cross-stack faults: SADE recovers the root cause; both
baselines fail. F = final outcome score, O = overall score.}
\label{tab:cases}
\begin{tabularx}{\linewidth}{@{}p{0.10\linewidth} >{\raggedright\arraybackslash}X >{\raggedright\arraybackslash}X@{}}
\toprule
 & \textbf{Case 1: \fp{bgp_acl_block}} & \textbf{Case 2: \fp{dns_port_blocked}} \\
\midrule
\textbf{Fault}   & \texttt{nft} on a spine drops TCP/179. & \texttt{nft} on the DNS pod filters TCP/UDP 53. \\
\textbf{GT}      & \fp{spine_router_2_3} & \fp{dns_pod2} \\
\midrule
\textbf{SADE}    & P1 clean (70/70 paths, 0\% loss) $\rightarrow$ \texttt{infra\_sweep} flags \fp{spine_router_2_3} $\rightarrow$ \texttt{acl-skill} confirms TCP/179 drop. 16 turns. F=5, O=5. & Pod reachability $\rightarrow$ direct port-53 probe $\rightarrow$ pod \texttt{nft} confirms \fp{dns_port_blocked}. F=5, O=5. \\
\textbf{ReAct}   & No router probe; submitted \fp{host_incorrect_dns} on \fp{client_0}. F=1, O=2. & Read mixed DNS as healthy; submitted \texttt{is\_anomaly=False}. F=1, O=2. \\
\textbf{CC-B}    & 21 turns of broad probes; never inspected a spine ACL. Budget exhausted. F=1, O=3. & Drifted to BGP control plane; submitted \fp{bgp_asn_misconfig}. F=1, O=2. \\
\midrule
\emph{Why ReAct} & Anchored on DNS, never falsified; no router probe, so the BGP-ACL fingerprint stays invisible. & Never probed the pod's port-53 path; partial DNS reads were read as a transient resolver issue. \\
\emph{Why CC-B}  & No termination rule; probing widened instead of narrowing, exhausting the budget before any spine \texttt{nft} ruleset was inspected. & Drifted to the BGP control plane after the DNS symptom and never returned to the pod. \\
\bottomrule
\end{tabularx}
\end{table}

\section{Skills Overview}
\label{app:skills_overview}

\begin{table}[H]
\centering\scriptsize
\renewcommand{\arraystretch}{1.15}
\setlength{\tabcolsep}{3pt}
\caption{SADE skills bank: one broad-search skill, two shared
utilities, and twelve fault-family skills.}
\label{tab:skills}
\begin{tabularx}{\linewidth}{@{}p{0.05\linewidth} p{0.30\linewidth} >{\raggedright\arraybackslash}X@{}}
\toprule
\textbf{Grp} & \textbf{Skill} & \textbf{Role / fingerprints} \\
\midrule
\multirow{3}{*}{\rotatebox[origin=c]{90}{\scriptsize Util.}}
  & \texttt{diagnosis-method}  & Drives Step~2 deep scan: L2 $\rightarrow$ control-plane $\rightarrow$ host $\rightarrow$ service. \\
  & \texttt{baseline-behavior} & Compares live signal against known-normal patterns; gates symptom commitment. \\
  & \texttt{big-return}        & Parses oversized tool output for diagnostically relevant fields. \\
\midrule
\multirow{12}{*}{\rotatebox[origin=c]{90}{\scriptsize Fault families}}
  & \texttt{link-fault}     & \fp{link_detach}, \fp{link_down}, \fp{link_flap} \\
  & \texttt{mac-conflict}   & \fp{mac_address_conflict} \\
  & \texttt{host-ip}        & \fp{host_ip_conflict}, \fp{host_wrong_ip}, \fp{host_wrong_gateway}, \fp{host_wrong_netmask}, \fp{host_missing_ip}, \fp{host_incorrect_dns}, \fp{host_static_arp} \\
  & \texttt{acl}            & \fp{arp_acl_block}, \fp{icmp_acl_block}, \fp{http_acl_block}, \fp{dns_port_blocked}, \fp{bgp_acl_block}, \fp{ospf_acl_block}, \fp{link_fragmentation_disabled} \\
  & \texttt{tc-fault}       & \fp{link_bandwidth_throttling}, \fp{link_high_packet_corruption}, \fp{incast_traffic_network_limitation} \\
  & \texttt{ospf-fault}     & \fp{ospf_neighbor_missing}, \fp{frr_service_down}, \fp{ospf_area_misconfiguration} \\
  & \texttt{bgp-fault}      & \fp{bgp_asn_misconfig}, \fp{bgp_missing_route_advertisement}, \fp{bgp_hijacking}, \fp{bgp_blackhole_route_leak}, \fp{host_static_blackhole} \\
  & \texttt{dhcp-fault}     & \fp{dhcp_service_down}, \fp{dhcp_missing_subnet}, \fp{dhcp_spoofed_subnet}, \fp{dhcp_spoofed_dns} \\
  & \texttt{dns-fault}      & \fp{dns_service_down}, \fp{dns_record_error}, \fp{dns_port_blocked}, \fp{dns_lookup_latency} \\
  & \texttt{load-balancer}  & \fp{load_balancer_overload} \\
  & \texttt{resource-cont.} & \fp{sender_resource_contention}, \fp{receiver_resource_contention}, \fp{sender_application_delay}, \fp{web_dos_attack} \\
  & \texttt{host-crash}     & \fp{host_crash} \\
\bottomrule
\end{tabularx}
\end{table}

The Skills Bank comprises 15 \texttt{SKILL.md} files in three groups,
summarised in Table~\ref{tab:skills}: a broad-search skill that drives
the Step~2 deep scan, two cross-cutting utilities consulted before
symptom commitment, and twelve fault-family skills invoked at Step~4
once a symptom has been mapped to its owning family. The
implementation \texttt{-skill} suffix is omitted from the cells.

\noindent\textbf{Diagnosis manual excerpt.} The diagnosis manual
defines the broad-search behavior used when the initial reachability
scan does not expose a confirmed symptom. In this case, the agent
enters the \texttt{diagnosis-methodology-skill} rather than committing
directly to a fault-family skill. The manual orders the deep scan into
four phases. Phase~A inspects L2 and infrastructure state using
\fp{infra_sweep.py} and \fp{l2_snapshot.py}; Phase~B
inspects routing, control-plane, and traffic-control state using
\fp{ospf_snapshot.py}, \fp{bgp_snapshot.py}, and
\fp{tc_snapshot.py}; Phase~C inspects host-local behavior using
\fp{host_path_snapshot.py}, \fp{dhcp_link_history.py}, and
\fp{safe_reachability.py}; and Phase~D inspects service and
resource-pressure behavior using \fp{service_snapshot.py},
\fp{pressure_sweep.py}, \fp{dns_client_snapshot.py}, and
\fp{http_client_snapshot.py}. The agent returns to the Fault
Index only after this broad search surfaces a confirmed symptom,
preventing speculative entry into a fault-family skill.

\vspace{12pt}

\end{document}